# Optical Gravity in a Graviton Spacetime

**Matthew R. Edwards**

John P. Robarts Library, 6th Floor, University of Toronto, Toronto, Ontario, Canada M5S 1A5

e-mail: matt.edwards@utoronto.ca

Orcid ID: 0000-0001-8817-458X

**Abstract**

While the Hubble redshift is generally linked to expansion of spacetime, an open question concerns where the energy lost from redshifted photons, gravitons and gravitational waves goes. One possibility is that it gives rise to gravity. In the optical-mechanical analogy in general relativity, relativistic light deflection is treated as refraction in an optical medium of varying density gradient. We now model the optical medium analogue of spacetime as a real graviton conjugate interlinking all masses. Gravitons in local mass systems are assumed to be organized in coherent quantum states overlapping with those of particles within spacetime structures. With the Hubble redshift, however, gravitons acquire longer wavelengths as they are shifted to less coherent states. Assuming that the momentum and energy lost from gravitons is transferrable to particles, mutual screening by masses then gives rise to an attractive force with a magnitude consistent with gravity. Some possible cosmological implications of the model are discussed.

**Keywords**: general relativity, quantum gravity, vacuum refraction index, graviton, Hubble redshift

# 1. Introduction

For all its great successes general relativity still stands apart from the world of quantum physics. It has yet to be shown reducible, for example, to the exchange of gravitons by masses in a successful theory of quantum gravity. Instead spacetime curvature is often depicted as being sufficient alone to dictate how matter moves under gravitation. One approach potentially linking general relativity and quantum theory resides in the optical-mechanical analogy in general relativity. Since the time of Eddington [1] it has been recognized that relativistic light deflection can also be modeled as a quasi-refraction of light through a medium with a density gradient and varying refraction index. In recent years this optical analogy has attracted increasing interest [2] [3] [4] [5] [6] [7] [8]. It has been used, for example, to make precise calculations of gravitational lensing [9] [10] [11]; to develop expressions for photons and particles moving in planetary orbits [4] [12]; to form a different model of the gravitational redshift [13] and in experimental 'toy' models even to simulate the trapping of light near the photosphere of supermassive astrophysical objects, such as black holes [14]. Building on this extremely precise correspondence between the process (relativistic light bending) and the analogy (refraction in an optical medium of varying density), some authors have further proposed that the optical medium is in fact a real polarizable medium and that it may even play a role in gravitation itself [6] [10] [15] [16] [17] [18]. In the model developed by Wilson and later Dicke, for example, gravity results from a vacuum refraction index that increases over time.

Using the Abraham expression for optical momentum, it was recently proposed by the author that gravity could arise with progressive loss of photon momentum and energy to a cosmic optical medium [17]. The argument hinged on the condition that the envelope of medium around a mass must be physically bound to that mass, such that envelope and mass can always move in concert even at relativistic speeds. In this case any momentum transferred from a photon to the envelope of optical medium about a mass would be ultimately transferred to the mass itself. A photon passing a mass and its medium envelope would be weakened in the process and would thus transfer less momentum and energy to a second mass encountered than it did to the first. Gravity might then result with masses effectively being pushed together by photons. This model of optical gravity was tested using the CMB, the largest available pool of electromagnetic energy known to exist. The CMB was found to be too weak to cause gravity, however, unless additional assumptions were made.

There is, however, an alternate energy field that in theory could be strong enough to drive gravity using the optical approach: gravitational energy itself. In general relativity, gravitational potential energy is associated with spacetime curvature, which is itself influenced by masses. As spacetime expands with the Hubble redshift, the gravitational potential around masses must seemingly diminish. While the fate of the energy lost from photons and the spacetime metric during expansion is typically disregarded (*e.g.*, [19]), a strict conservation of energy and matter can nonetheless be contemplated. If so, the lost energy could conceivably be redirected into gravitational attraction. Spacetime curvature then would not directly guide the motion of matter but would instead move it indirectly through the energy lost from it. Similarly, in the quantum gravity approach, the Hubble expansion of spacetime might similarly diminish the energies of embedded gravitons, with the lost energy again being redirected to gravity.

In this context it is noteworthy that much evidence already exists of gravitational potential energy being converted to other forms of energy under the Hubble redshift. Numerous findings suggest that the internal gravitational potential energies of astrophysical and geophysical objects, including stars, planets and supermassive black holes, decays according to the relationship $L_H = -UH_0$, where $U$ is the object's internal gravitational potential energy, $H_0$ is the Hubble constant and $L_H$ is the object's 'Hubble luminosity' [20] [21] [22]. Here the mutual gravitational potential energy of masses is assumed to be equal in magnitude to the total graviton energy being continuously exchanged between them. Significantly, if gravitons were organized in graviton filaments connecting all masses, as previously suggested [20] [21] [22], the cosmic network so derived could serve as the cosmic optical medium required for optical gravity. The firm linkages between graviton filaments and masses would satisfy the criterion in optical gravity that the envelopes of optical medium about masses are attached to and so can move together with the host masses.

But if gravitons are to serve as the cosmic optical medium in optical gravity, can they also substitute for photons as the driving agent? The only additional assumption we would need to make is that a graviton in traversing a graviton optical medium is refracted in the same way as any other wave. This can again be easily visualized if gravitons are arranged in semi-solid structures interconnecting all masses. The energy and momentum lost from a graviton by refraction would then simply be transferred to other gravitons or gravitational waves (GWs) moving in different directions. Over time this refraction and re-refraction of gravitons would produce a uniform, low-energy graviton background analogous to the CMB. Gravity would then proceed as in our earlier photon model. In this paper we consider this graviton model in detail, beginning by reviewing and updating our earlier treatment of optical gravity.

## 2. Photon Refraction in a Cosmic Optical Material

Optical gravity hinges on the expression for optical momentum in an optical material. While Minkowski originally proposed that a light pulse should have momentum $nE/c$ in a dielectric medium of refractive index *n*, Abraham used a different approach to argue that it should be *E/(nc)*. Ever since this issue has been the subject of a controversy still not close to being settled [23] [24] [25] [26] [27] [28] [29] [30] [31].

In a simple thought experiment, however, Balazs [32] used arguments of momentum conservation to show that a photon with initial energy $\hbar\omega$ and speed *c* before entering a transparent block can only have the Abraham momentum $p = \hbar\omega/cn$ once inside the block (for recent discussions see [24] [33] [34]). In this case the block must acquire the momentum

$$p_{bl} = \left(1 - \frac{1}{n}\right)\frac{\hbar\omega}{c} \tag{1}$$

while the photon is inside it. Note that a similar result would be obtained if the optical medium is weakly absorbing, since the rate of photon absorption at local (*e.g.*, intra-galactic) scales would be miniscule (see [24]). In their analysis of the Balazs experiment in various media types, Chau and Lezec [33] concluded that the conservation laws support the Abraham momentum density as the true electromagnetic momentum density.

Using the Abraham interpretation, the loss of a photon's momentum to a transparent block of optical medium associated with a particular mass was estimated as follows [17]. For low potentials, *i.e.*, where $GM/rc^2 \ll 1$, the index of refraction *n* in the optical-mechanical approach can be approximated as

$$n \cong 1 + \frac{2GM}{rc^2} \tag{2}$$

[10]. From (1) and (2) the momentum acquired by the block of medium from the photon is then

$$p_{bl} \cong p\frac{2GM}{rc^2}. \tag{3}$$

Spacetime would then consist of the overlapping blocks of optical medium associated with all the masses in the visible universe. Neglecting possible effects due to universal expansion, the index of refraction of the visible universe was estimated as

$$n_U = 1 + \frac{2G}{c^2} \int_0^{R_U} \frac{4\pi\rho r^2}{r} dr. \tag{4}$$

Since $R_U = c/H_0$, we then have

$$n_U = 1 + \frac{4\pi G\rho}{H_0^2}. \tag{5}$$

From (3), (4) and (5), it can be seen that the momentum transferred from a photon to the cosmic medium while it travels the whole distance $R_U$ would be $p(4\pi G\rho/H_0^2)$, where $p$ is the initial photon momentum. This momentum transfer would occur over a time interval equal to the Hubble time $H_t = 1/H_0$. For a photon situated at the origin of the cosmic sphere and travelling a short distance $dl = c\, dt$, the rate of momentum transfer as it travels the small distance $dl$ is then given by

$$dp = p\left(\frac{4\pi G\rho}{H_0^2}\right)\left(\frac{dl}{R_U}\right). \tag{6}$$

Since $dl = c\, dt$ and with $R_U = c/H_0$, the instantaneous rate at which momentum is transferred to spacetime is then

$$\frac{dp}{dt} = p\left(\frac{4\pi G\rho}{H_0}\right). \tag{7}$$

Taking $\rho = 5 \times 10^{-30}$ gm cm$^{-3}$, which is about equal to the critical density $\rho_c \equiv 3H_0^2/8\pi G$, the photon correspondingly loses momentum to spacetime at the fractional rate

$$\frac{\dot{p}}{p} = -2 \times 10^{-18} s^{-1} = -H_0. \tag{8}$$

Similarly, for the fractional rate of loss of photon energy $E$, we would have

$$\frac{\dot{E}}{E} = -H_0. \tag{9}$$

As these rates are close to those experienced by light due to the Hubble redshift, it was suggested that they could account for at least part of the redshift.

As described earlier, the loss of photon energy and momentum to a cosmic optical medium was then used to derive a simple gravity model [17]. A photon in its trajectory through space would contact the envelope of optical medium around each individual mass separately and lose momentum and energy to it. Since each mass-centred envelope of medium contacted would acquire less momentum from the photon than the one before, the two masses would be pushed together. Photons were found to have insufficient energy to drive gravity, however, unless additional assumptions concerning the role of dark matter were made.

## 3. Spacetime as a Graviton Conjugate

The alternative source of energy now suggested to drive optical gravity is a graviton background associated both with the universe's total gravitational energy and with spacetime curvature. Such a background would be consistent with the notion that the GWs propagating in spacetime are themselves comprised of gravitons. The magnitude of graviton energy in a gravitating system is again assumed to be equal to the gravitational potential energy $U$ of that system. In this connection, it has been suggested that gravitons might find their densest concentrations in black holes, where they possibly form a quasi-Einstein-Bose condensate [35] and their thinnest one in a graviton conjugate filling all space, by virtue of the gravitational interactions of a mass with all the distant stars of the visible universe [36]. As noted above, the specification of fixed linkages of graviton filaments to masses is key to the operation of optical gravity: it allows optical momentum transferred to filaments to be passed on to the host masses. A graviton conjugate so constructed would also have its highest density where filaments converge near masses, thus satisfying the requirement in the optical-mechanical analogy that the index of refraction and density of the medium at a point around a mass are proportional to the gravitational field strength at that point. Gravitons modeled as virtual photons of some sort would also connect generally with the polarizable vacuum approach [6] [17] [21]. They might then be seen as maintaining spacetime structures in like fashion to the virtual photons which maintain atomic electronic structures.

A graviton conjugate interlinking all masses is also compatible with a central tenet of general relativity: that spacetime must continuously be updated at each point in space by the masses embedded within it. In a graviton spacetime, information concerning a particle (*e.g.*, its position, velocity or spin) must be continuously carried away from it by gravitons, presumably at the speed of light. In each graviton-particle interaction the graviton's configuration must therefore be altered in such a way that it encodes updated information about the particle. Particles in some sense must thus act as reprocessing centres for gravitons, converting gravitons with older, outdated information originating from remote masses into newer ones carrying updated information about local particles.

## 4. Optical Gravity with Gravitons

The model of optical gravity being proposed here features gravitons as both the optical medium and as the driving agent, substituting for photons. As energy quanta embedded in spacetime, gravitons would be subject to relativistic refraction within other graviton structures they encounter during their travels, with a concomitant Hubble redshift and loss of energy and momentum. Accordingly, and noting that $U$ is negative, the expression for the loss of graviton energy within a mass or mass system with internal gravitational potential energy $U$ would be

$$\frac{dE}{dt} = -UH_0. \tag{10}$$

Within large masses like the Earth, the energy and momentum delivered to the component atoms due to this process would be significant, as $U$ in these situations is generally proportional to $M^2$. The release of energy inside planets, neutron stars, white dwarfs and supermassive black holes from (10) would be enormous and has been suggested to account for hitherto unexplained excess luminosities and/or tectonic processes in these objects [20] [21] [22]. It is this energy released from gravitons that drives gravity in our model.

The gravitons of special interest for optical gravity would be the ones a mass exchanges with the most distant masses in the visible universe. As noted by Tryon [37] and others, most of the gravitational potential energy associated with a mass, $E_g$, resides in these most remote interactions. Since gravitons would be weakened by the Hubble redshift in the same way as photons, it can be inferred that gravity would fall off exponentially with distance. This effect can be incorporated in the expression for $E_g$ by adding an effective extinction coefficient term $\alpha$ along the path between two masses, where $\alpha = H_0/c$ [38].

If matter, chiefly in the form of galaxies, is considered to be distributed approximately evenly in space, $E_g$ can then be evaluated by integrating the gravitational potential energy of concentric shells of matter of thickness $dr$ centred about $m$, each of constant density $\rho$, out to infinite distance. We then have

$$E_g = -\int_0^\infty \frac{4\pi G\rho m r^2 e^{-\alpha r}}{r} \, dr = -4\pi G\rho m \int_0^\infty r e^{-\alpha r} dr. \tag{11}$$

Substituting from above for $\alpha$, we then have

$$E_g = -\frac{4\pi G\rho m c^2}{H_0^2}. \tag{12}$$

For $\rho$ equal to the critical density $\rho_c = 3H^2/8\pi G$ we find

$$E_g = -mc^2, \tag{13}$$

*i.e.*, about twice the value estimated by Tryon. Note that $E_g$ has a finite value even if the universe is infinite in extent. The relationship in (13) formed the basis for Tryon's speculation that the universe arose 'out of nowhere' in a vacuum fluctuation, which has since undergone many elaborations (see [39]). In Tryon's model the gravitational potential energy of newly formed matter cancels the rest energy of that matter, thus retaining universal conservation of energy. In general, this idea meshed with Mach's notion that the inertial mass of a body rests in its interactions with the distant stars.

The graviton energy lost from $E_g$ due to the Hubble redshift is eventually absorbed by all the masses of the visible universe. In this process, discussed more fully below, redshifted gravitons are converted by particles to updated, non-redshifted gravitons. From (10) and (13) the rate at which a mass absorbs graviton energy due to the Hubble redshift is then

$$\frac{dE}{dt} = mc^2 H_0. \tag{14}$$

Gravity then arises simply as follows (Fig. 1). Due to the uniform distribution of remote masses in space, $M_1$ receives momentum boosts from gravitons symmetrically and so no net force arises. If a second mass, $M_2$, is introduced at a distance $d$ from $M_1$, where $d \ll R_U$, however, some of the gravitons that would otherwise have been received by $M_1$ are now intercepted by $M_2$. From the opposite direction, the flux of gravitons striking $M_1$ is undiminished in strength, however, and so $M_1$ experiences a net acceleration towards $M_2$. For the same reasons, $M_2$ experiences an equal and opposite push towards $M_1$. The net result is that the two masses are pushed together.

Only the Hubble redshift is then needed to quantify the gravitational force. As before, the linear absorption coefficient of graviton energy is determined from the Hubble redshift to be $\alpha = H_0/c$. From $\alpha$ we then derive the effective mass absorption coefficient $h$ as

$$h = \frac{\alpha}{\rho} = \frac{H_0}{\rho c}, \tag{15}$$

with units of cm$^2$ gm$^{-1}$ [17]. For $H_0$ = 66 km s$^{-1}$ Mpc$^{-1}$ = 2.2 × 10$^{-18}$ s$^{-1}$ gm$^{-1}$ and for $\rho$ = 5 × 10$^{-30}$ gm cm$^{-3}$, we then have $h \cong$ 10 cm$^2$ gm$^{-1}$. This large value for $h$ compared to values determined theoretically and in gravitational shielding experiments simply reflects that the initial removal of graviton energy from its original graviton structure does not occur initially in the nuclear subunits of masses, but rather in their associated envelopes of optical medium extending outwards from the masses to cosmic distances.

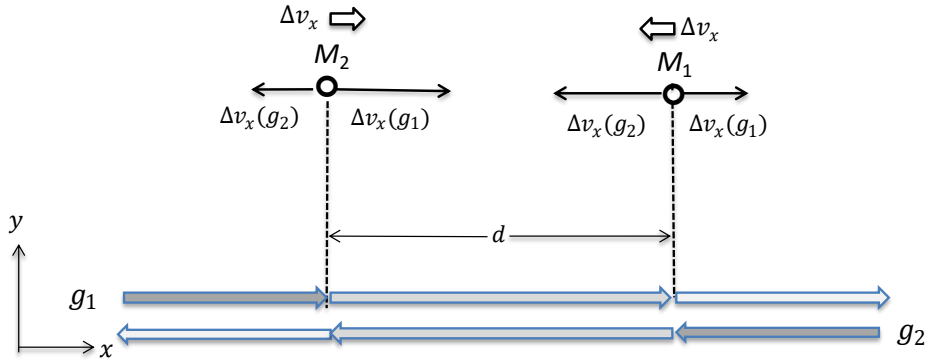

**Fig. 1.** Optical gravity via refraction in a graviton spacetime. Two gravitons originating from remote masses enter from the left and right. As each one passes one of the masses, it gives up part of its momentum and energy first to the envelope of optical medium (spacetime) fixed to that mass and then to the mass itself. Each graviton will now have a reduced momentum and energy and so will transfer less of those to the second mass it encounters. The second mass will accordingly be pushed less than the first, producing a net motion of the two masses towards each other.

To derive the force on $M_1$ we then only require the fraction of graviton energy ordinarily absorbed by $M_1$ that is now intercepted by $M_2$. This fraction is given by $M_2 h/4\pi d^2$. The missing energy flux on $M_1$ due to the presence of $M_2$ is then

$$\frac{dE}{dt} = - M_1 c^2 H_0 \times \frac{M_2 H_0}{4\pi c \rho d^2}.\qquad(16)$$

For graviton momentum $E/c$, the resulting force acting on $M_1$ in the direction of $M_2$ is then

$$F = - \frac{M_1 M_2 H_0^2}{4\pi \rho d^2}.\qquad(17)$$

From this it can be seen that the term for the gravitational constant is

$$G = \frac{H_0^2}{4\pi \rho}.\qquad(18)$$

For $\rho = \rho_c = 5 \times 10^{-30}$ gm cm$^{-3}$ and $H_0$ as before, we find $G = 7 \times 10^{-8}$ cm$^3$ gm$^{-1}$ s$^{-2}$, which is very close to its actual value of $6.67 \times 10^{-8}$ cm$^3$ gm$^{-1}$ s$^{-2}$. Thus, for $\rho \cong \rho_c$, the model is able to recover Newton's law.

The above approach is similar to versions of Le Sage's theory of gravity in which a cosmic field of electromagnetic radiation gives rise to the force, such as that of Radzievskii and Kagalnikova [40].

A key feature of Le Sage models is that the gravitational flux must be absorbed by masses, not merely reflected, or else the shading effect is entirely nullified by rebounding particles or waves [41]. To account for this in the model it is necessary to consider more closely the processing of spacetime by matter. As noted above, a key tenet of general relativity is that masses continuously update spacetime, which in the present case is in the form of filaments of gravitons connecting all masses. Due to this updating spacetime structures enclosing local mass systems such as galaxies would be the strongest, formed of essentially non-redshifted, coherently organized gravitons. With spacetime expansion, however, these structures are weakened, with gravitons gradually acquiring longer wavelengths and progressively incoherent states. Upon subsequent absorption and updating by other matter, the disordered and weakened gravitons would be reorganized into new, coherently organized spacetime elements. The disordered gravitons would transfer energy and momentum to particles in the process, but the gravitons in outgoing, newly refurbished spacetime elements would not. Like the electrons in an atom the quantum states of local spacetime structures and their enclosed masses would coherently overlap to maintain system stability.

Redshifted gravitons would, however, induce repulsive forces between masses. For local systems, these forces can be shown to be negligible in comparison to the gravitational force. Let us suppose that in a system of two masses the product waves of graviton decay travel back along the axis connecting the masses and strike each mass. The repulsive force between the two masses is then

$$F = \frac{Gm_1 m_2 H_0}{rc}. \tag{19}$$

The factor $c$ appears in the denominator since a quantum of radiation with energy $E$ has momentum $E/c$. In the Earth-Moon system, for example, the repulsive force on the Moon would result in an acceleration of only $\sim 10^{-18}$ cm s$^{-2}$, which is entirely negligible in comparison to the gravitational attraction. The two forces are seen to balance only when $r \approx R_U \approx c/H$.

## 5. Spacetime Expansion: with or without Cosmic Expansion?

Previously, it was noted that a redshift effect linked to refraction of light in a cosmic optical medium could be consistent either with an expanding universe or a static one and that distinguishing between these two possibilities could be a difficult task [17]. In this light consider the observed time dilation in supernovae, long seen as one of the strongest pieces of evidence supporting an expanding universe. We have conjectured that spacetime is itself made up of gravitons and that these lose momentum and energy in moving through other pockets of spacetime in the same manner as photons. In that event, individual packets of spacetime would be subject to the Hubble redshift *even in the absence of cosmic expansion*.

Consider a cluster of photons emitted from an exploding supernova. In either BBT or the present scheme the photons are embedded within spacetime. In BBT, time dilation in the supernova arises due to expansion of spacetime since the time of the explosion. The light curve of the distant supernova is stretched in the observer's frame by a factor $(1 + z)$ compared to the supernova's rest frame [42]. In the present model, the photons occupy a particular spacetime packet for a very short time interval $dt$. But during that interval, the spacetime packet would also lose energy due to the model redshift process. Its constituent gravitons would be redshifted to longer wavelengths and the packet as a whole would thus expand a tiny amount during the time interval $dt$. Once the photon cluster has passed into the adjacent packet of spacetime, the distance between two photons within this packet will accordingly have increased. This effect would produce the same time dilation effect as in BBT, but without universal expansion. As noted above, spacetime expansion and thinning would be balanced by spacetime rejuvenation with the absorption and reemission of gravitons at mass centres. The evidence of time dilation in supernovae thus does not unequivocally support the expanding universe cosmology. It could also allow for a static universe involving recycling of all constituents.

One of those constituents is the CMB. The CMB is reasonably well explained in BBT, wherein the smooth blackbody spectrum is preserved while the CMB cools with spacetime expansion. Yet, as already noted, this transition corresponds to a steady loss of energy from the universe if no accounting is made of the lost CMB energy. For example, CMB temperatures in earlier epochs, when the universe was smaller in BBT, are thought to be elevated compared to that of the CMB today. Backing this are studies reporting excitation of molecules in remote gas clouds billions of years ago. Against this, however, the present model allows for a different mechanism to explain the heating of those high-$z$ gas clouds. By virtue of their enormous masses those clouds possess considerable gravitational potential energy of their own, and thus through (10) would be subject to an internal heating effect.

While the existence and approximate magnitude of the CMB had been predicted in static models prior to the first detection of the CMB in the 1960s, it has proved very difficult to square the CMB spectrum with models not based on universal expansion. Here again a mechanism for producing and maintaining the CMB without cosmic expansion can be envisaged in optical gravity. As gravitons are jostled out of highly ordered states by the Hubble redshift, they would acquire weaker, more random states characterized by longer wavelengths. Since photons are likewise embedded in the spacetime fabric, it would follow by analogy that the energy they lose via the redshift goes into random electromagnetic waves of very long wavelength. These random waves would be generated uniformly in all parts of space and would heat up interstellar gas to a uniform temperature, which in turn would eventually give rise to the CMB with its characteristic blackbody spectrum.

Long ago it was recognized by Kelvin that an unending conversion of the universe's mechanical energy to thermal energy would steadily increase the entropy of the universe, ultimately culminating in its 'heat death'. This was considered a natural outcome of the second law of thermodynamics. In the process of CMB formation suggested above, however, we see that highly redshifted photons weaker than CMB photons are reorganized by matter to create the smooth CMB background.

If such a picture were true, then the notion that the universe must eventually suffer a 'heat death' due to steadily increasing entropy would be overturned. Graviton and photon energy would be steadily lost to the cosmic optical medium due to refraction, but the lost energy would then be rechanneled back into higher energy gravitons and photons. The universe in this case might then be better viewed, as it was in the past, as a cosmic *perpetuum mobile*, with all energy forms being continuously recycled.

In this connection, it is worthwhile to touch on the issue of apparent loss of information when an object falls into a black hole. Optical gravity again allows for a different interpretation here. A particle in the outer shell of a black hole would exist between the two worlds – an external and an internal one – with graviton filaments connecting it to both. Whether a particle exists in this shell, inside the black hole core or exterior to the black hole, however, we infer from our model that the total energy of the graviton filaments connected to it is always equal to its rest energy $m_0c^2$. The particles of the black hole core would thus always be connected to each other by graviton filaments similarly as free particles. Any information deposited in the surface shell could thus potentially pass into the black hole's internal spacetime, reappear within the shell or be reradiated to space in the reprocessed gravitons of the shell. The information in particles falling into a black hole thus would not at any time be lost from the universe.

## 6. Concluding Remarks

In this paper we have shown that if spacetime is construed as a graviton optical medium, with total energy equal to the universe's store of gravitational potential energy, then the energy lost from it due to the Hubble redshift is sufficient to drive gravitation. The gravitons of spacetime were suggested to be organized coherently in filaments interconnecting all masses. Under the Hubble redshift these filaments become weakened, with the lost energy going into incoherent quantum states. Reprocessing of redshifted gravitons by matter back into coherent forms then permits a simple gravity model based on mutual screening of gravitons by masses.

It might be argued that the gravitons exchanged by particles are just that – gravitons – as opposed to newly formed spacetime. But there is a fundamental consideration to be made here. If the energy contained in gravitons is sufficient to drive the most basic processes in physics, including gravity, is there any valid reason to suppose the existence of an *extra* energy pool or entity which comprises spacetime itself? Surely it would be more economical to make this identification: that spacetime consists of no more than the tight lattices of graviton beams generated by and connecting all the particles in the visible universe.

The model scheme potentially allows some degree of reconciliation between the views of gravity espoused in general relativity and theories of quantum gravity. Spacetime and spacetime curvature would be formed of graviton filaments, but spacetime curvature would not cause gravity directly. Gravity would instead be caused by decoherence and weakening of gravitons induced by the Hubble redshift. The model is consistent either with an expanding universe of decreasing energy density or a non-expanding universe of constant energy density, where the Hubble redshift has a different cause than cosmic expansion.

The approach we have taken could lastly have some major implications for quantum theory. A graviton spacetime would seemingly furnish a basis for quantum entanglement, as every particle is linked to every other particle through graviton filaments. Any two particles which closely interact with each other would thus carry the record of their interaction until such time as their quantum states are disturbed in subsequent interactions. It would also seem to mesh with the de Broglie-Bohm or 'hidden variables' approach in quantum physics, since matter waves, for example, might be modeled as modulations of graviton carrier waves induced by matter in motion.

**Conflict of interest statement**

The author has no relevant financial or non-financial interests to disclose.

**Data availability statement**

Data sharing not applicable to this article as no datasets were generated or analysed during the current study.